\documentclass[11pt]{article}

\usepackage{graphicx}
\usepackage{amsmath}
\usepackage{amssymb}
\numberwithin{equation}{section}

\newcommand{\fL}{{\mathfrak L}}
\newcommand{\fR}{{\mathfrak R}}
\newcommand{\fB}{{\mathfrak B}}
\newcommand{\fF}{{\mathfrak F}}
\newcommand{\fQ}{{\mathfrak Q}}
\newcommand{\fS}{{\mathfrak S}}
\newcommand{\fC}{{\mathfrak C}}

\newcommand{\fP}{{\mathfrak P}}
\newcommand{\fK}{{\mathfrak K}}

\newcommand{\fJ}{{\mathfrak J}}
\newcommand{\fU}{{\mathfrak U}}

\newcommand{\alg}[1]{\mathfrak{#1}}

\newcommand{\comm}[2]{[#1,#2]}
\newcommand{\gen}[1]{\mathfrak{#1}}
\newcommand{\cwgen}[1]{#1}
\newcommand{\psu}{\alg{psu(2|2)}}
\newcommand{\psucentral}{\alg{psu(2|2)}\ltimes\mathbb{R}^3}
\newcommand{\Ya}[1]{\mathcal{Y}(#1)}
\def\[{\begin{equation}}
\def\]{\end{equation}}
\def\<{\begin{eqnarray}}
\def\>{\end{eqnarray}}

\newcommand{\half}{\frac{1}{2}}

\newcommand{\adj}[3]{\left(\text{ad}_{#1}\right)^{#2}(#3)}
\newcommand{\adjb}[2]{\text{ad}_{#1}(#2)}
\newcommand{\adjbb}[1]{\text{ad}_{#1}}
\newcommand{\acomm}[2]{\{#1,#2\}}
\newcommand{\copro}{\Delta}
\newcommand{\eip}{\mathcal{U}}
\newcommand{\nln}{\nonumber\\}
\newcommand{\nn}{\nonumber}
\newcommand{\genY}[1]{\widehat{\mathfrak{#1}}}
\newcommand{\nl}[1][0pt]{\nonumber\\[#1]&\hspace{-4\arraycolsep}&\mathord{}}
\newcommand{\quarter}{\frac{1}{4}}
\newcommand{\grp}[1]{\mathrm{#1}}
\newcommand{\struc}{f}
\newcommand{\earel}[1]{\mathrel{}&\hspace{-2\arraycolsep}#1\hspace{-2\arraycolsep}&\mathrel{}}
\newcommand{\eq}{\earel{=}}

\begin{document}

\baselineskip=16pt plus 0.2pt minus 0.1pt
\begin{titlepage}
\begin{flushright} 
MIT-CTP 3935\\
Imperial/TP/08/FS/01\\
HU-EP-08/05\\
\end{flushright}
\mbox{ }  \hfill 
\vspace{5ex}
\Large
\begin {center}     
{\bf On Drinfeld's second realization of the AdS/CFT $\alg{su(2|2)}$ Yangian}
\end {center}
\large
\vspace{1ex}
\begin{center}
Fabian Spill ${}^a$ \ and \ Alessandro Torrielli ${}^b$
\end{center}
\vspace{1ex}
\begin{center}
${}^a$\textit{Blackett Laboratory,
Imperial College London\\
London SW7 2AZ, UK\\
and\\
Humboldt-Universit\"at zu Berlin, Institut f\"ur Physik,\\%
Newtonstra\ss{}e 15, D-12489 Berlin, Germany}\vspace{3mm}\\
\texttt{fabian.spill@imperial.ac.uk}\\
\vspace{2ex}
${}^b$ \textit{Center for Theoretical Physics\\
Laboratory for Nuclear Sciences\\
and\\
Department of Physics\\
Massachusetts Institute of Technology\\
Cambridge, Massachusetts 02139, USA}\\ 
\vspace{1ex}
\texttt{torriell@mit.edu}

\end{center}
\vspace{4ex}
\rm
\begin{center}
{\bf Abstract}
\end{center} 

We construct Drinfeld's second realization of the Yangian based on $\psucentral$ symmetry. The second realization is traditionally more suitable for deriving the quantum double and the universal R-matrix
with respect to the first realization, originally obtained by Beisert, and it is generically more useful in order to study finite dimensional representations. We show that the two realizations are isomorphic, where the isomorphism is almost the standard one given by Drinfeld for simple Lie algebras, but needs some crucial corrections to account for the central charges. We also evaluate the generators of the second realization on the fundamental representation, finding the interesting result that the rapidity variable for some generators gets boosted by the energy eigenvalue.

\normalsize 

\vfill
\end{titlepage} 

\section{Introduction}
A remarkable recent development in the study of the AdS/CFT 
conjecture is
due to the observation that the dilatation operator of 
4D ${\cal{N}}=4$ supersymmetric Yang-Mills theory can be mapped
into the Hamiltonian of an integrable spin-chain \cite{MZ,gauge}. 
This matches an 
analogous integrable structure discovered on the string theory side of the 
correspondence \cite{string,nolocalo}. 
Integrability implies
that the relevant dynamical information is encoded in
the two-body spin-chain 
S-matrix. The tensor part of the latter is completely determined by the underlying centrally extended $\alg{su(2|2)}$ Lie superalgebra symmetry
in its fundamental representation \cite{B}, and satisfies the quantum Yang-Baxter equation \cite{nlin,Gleb}.
The S-matrix has to be supplemented
with an overall interpolating dressing factor, which is constrained, but not fully fixed, by a crossing equation derived in \cite{J}. The appearance of crossing symmetry is related to the existence of an underlying Hopf algebra, which was found in \cite{GH,PST}. This was confirmed by a string sigma model computation in \cite{KMRZ}, building upon the representation of \cite{nolocalo}. 
A remarkable 
non-perturbative solution of the crossing equation of \cite{J} has been proposed in \cite{BHLBES}, so far receiving highly non trivial 
confirmations (see for example \cite{BCDKS}).    

Several indications suggest the presence of a rich hidden mathematical structure responsible for the integrability
of the model (see also \cite{Hubbard}, and the recent treatment of a quantum 
deformation in \cite{NP}). To completely unravel it, it is important to understand how to embed the $\alg{su(2|2)}$ Lie
superalgebra in an infinite dimensional non-abelian (Yangian) symmetry \cite{Yangian}, and to construct the
so-called universal ({\it i.e.} representation-independent) R-matrix (see for instance \cite{Etingof}). Specifying such universal tensor gives immediate access to all possible scattering matrices in the various representations, as for instance it appears to be desirable from recent considerations on the TBA \cite{Glebult}. In fact, a better understanding of the structure of the asymptotic S-matrix would help progressing the finite-size problem as well \cite{finite}, since current approaches reutilizes the asymptotic data in order to draw conclusions on the compact case.

In \cite{BYang}, the S-matrix has been shown to possess a centrally extended $\alg{su(2|2)}$ Yangian symmetry. Hence, it is natural to expect such S-matrix to be the representation of the universal R-matrix of a Yangian double \cite{Drin}, somehow modified by a twist in order to incorporate the braiding elements of \cite{GH,PST}. Universal R-matrices for Yangians based on simple Lie (super)algebras have been derived in \cite{KT,stuko}, but a main ingredient in the derivation, the existence of a non-degenerate invariant bilinear form, is missing for $\psucentral$. One can in principle remove this degeneracy by adjoing the external $\alg{sl(2)}$ automorphisms to $\psucentral$, but they have no finite dimensional representation, and they cannot directly be seen to be symmetries of the S-matrix. 

As there exist no simple solution to overcome this problem, in \cite{T} a classical limit of the S-matrix and the centrally extended $\psucentral$ Lie algebra was studied. The
residue of the classical r-matrix was shown to possess an enhanced 
$\alg{u(2|2)}$ 
symmetry. Indeed, in \cite{BS} it was shown that this classical r-matrix on the fundamental representation arises from the abstract r-matrix of a quasitriangular bialgebra based on a deformed $\alg{u(2|2)}$ loop algebra (see \cite{MT} for an alternative proposal). This loop algebra contains an infinite tower of automorphisms, which one should in principle also expect to appear at the quantum level as symmetries of the S-matrix. However, to the present date only one additional Yangian generator, of $\alg{u(2|2)}$ signature, which can be interpreted as one of these automorphisms, was shown to be a symmetry \cite{MMT,BS}, whereas the role of the infinitely many other automorphisms remains to be uncovered.

Solving the problem of the singular Cartan matrix is not the only task one has to undertake in order to construct the universal R-matrix. The derivation performed in the literature \cite{KT} makes use of the so-called Drinfeld's second realization of the Yangian, which defines the Yangian in a Chevalley-Serre type of basis. In the case of simple Lie algebras Drinfeld also gave the isomorphism between the first and the second realization \cite{Dsecond}. To obtain the latter is the scope of the present paper. 

In this work, we define the Yangian $\Ya{\psucentral}$ based on the centrally extended $\psucentral$ algebra in Drinfeld's second realization, and show that it is isomorphic to the first realization of Beisert \cite{BYang}. We find that the defining relations are similar to those for Yangians of simple Lie algebras, but one needs to modify the Serre relations. Additionally, the isomorphism we find is similar to the case of simple Lie algebras, but needs some crucial modifications to incorporate the effects of the central elements.

Finally, we study the fundamental evaluation representation of the Yangian generators in the second realization. In the first realization all generators $\genY{J}^A$ at first level are represented as $\genY{J}^A \propto u\gen{J}^A$, namely by the level zero generators multiplied by a rapidity variable. In the second realization we find the interesting result that for some generators the rapidity variable gets shifted by the eigenvalue of the central charge $\gen{C}$, corresponding to the excitation energy, yielding effective ``boosted'' rapidities
\begin{eqnarray}
\omega_1 = i g u = \frac{i E}{\sin[p/2]} \cos[p/2], \qquad \omega_2 = i g u - C = \frac{i E}{\sin[p/2]} \exp[ip/2].
\end{eqnarray}
On the one hand, this is due to the presence of the threefold central extension, on the other hand its unusual character  further complicates the problem of constructing the universal R-matrix with standard methods, and we leave this issue for future investigations. We will nevertheless find a sensible ``triangular'' decomposition of the Yangian algebra, a nice realization in terms of Drinfeld currents, and a quite non trivial consistency check of this structure furnished by the Serre relations.

\subsection{Summary of the results}\label{summary} 
We summarize here the results we have obtained. The centrally extended 
$\alg{su(2|2)}$ part of the Yangian algebra
underlying the AdS/CFT S-matrix \cite{B} can be cast into the following 
set of defining relations for the fermionic roots $\xi^\pm_{i,n}$
and Cartan generators $\kappa_{i,n}$, with $i=1,2,3$ and $n=0,1,\dots$: 
\begin{align}
\label{relazionizero}
&[\kappa_{i,m},\kappa_{j,n}]=0,\quad [\kappa_{i,0},\xi^+_{j,m}]=a_{ij} \,\xi^+_{j,m},\nonumber\\
&[\kappa_{i,0},\xi^-_{j,m}]=- a_{ij} \,\xi^-_{j,m},\quad \{\xi^+_{i,m},\xi^-_{j,n}\}=\delta_{i,j}\, \kappa_{j,n+m},\nonumber\\
&[\kappa_{i,m+1},\xi^+_{j,n}]-[\kappa_{i,m},\xi^+_{j,n+1}] = \frac{1}{2} a_{ij} \{\kappa_{i,m},\xi^+_{j,n}\},\nonumber\\
&[\kappa_{i,m+1},\xi^-_{j,n}]-[\kappa_{i,m},\xi^-_{j,n+1}] = - \frac{1}{2} a_{ij} \{\kappa_{i,m},\xi^-_{j,n}\},\nonumber\\
&\{\xi^+_{i,m+1},\xi^+_{j,n}\}-\{\xi^+_{i,m},\xi^+_{j,n+1}\} = \frac{1}{2} a_{ij} [\xi^+_{i,m},\xi^+_{j,n}],\nonumber\\
&\{\xi^-_{i,m+1},\xi^-_{j,n}\}-\{\xi^-_{i,m},\xi^-_{j,n+1}\} = - \frac{1}{2} a_{ij} [\xi^-_{i,m},\xi^-_{j,n}],
\end{align}
\begin{eqnarray}
&&i\neq j, \, \, \, \, \, n_{ij}=1+|a_{ij}|,\, \, \, \, \, Sym_{\{k\}} [\xi^+_{i,k_1},[\xi^+_{i,k_2},\dots \{\xi^+_{i,k_{n_{ij}}}, \xi^+_{j,l}\}\dots\}\}=0,\nonumber\\
&&i\neq j, \, \, \, \, \, n_{ij}=1+|a_{ij}|,\, \, \, \, \, Sym_{\{k\}} [\xi^-_{i,k_1},[\xi^-_{i,k_2},\dots \{\xi^-_{i,k_{n_{ij}}}, \xi^-_{j,l}\}\dots\}\}=0,\nonumber\\
&&\text{except for} \, \, \, \, \, \, \, \, \, \{\xi^+_{2,n},\xi^+_{3,m}\} = \alg{K}_{n+m}, \qquad  \{\xi^-_{2,n},\xi^-_{3,m}\} = \alg{P}_{n+m}, 
\end{eqnarray}
where the symmetric 
Cartan matrix $a_{ij}$ has all zeroes except for $a_{12}=a_{21}=-1$ and $a_{13}=a_{31}=1$. The fundamental evaluation representation is given by
\begin{eqnarray}
\label{mult1zero}
&&\kappa_{i,n} = {\omega_i}^n \kappa_{i,0},\quad \xi^+_{i,n} = {\omega_i}^n \, \xi^+_{i,0},\quad \xi^-_{i,n} = {\omega_i}^n \, \xi^-_{i,0},\nonumber \\
&&\alg{K}_{n} = {\omega_2}^n \alg{K}, \qquad \alg{P}_{n} = {\omega_2}^n 
\alg{P},
\end{eqnarray} 
\begin{align}
\label{mult2zero}
\omega_1= ig u,\qquad \, \, 
\omega_2 = \omega_3 = ig u - C, 
\end{align}
$C$ being the eigenvalue of the central charge $\fC$.
The coproducts, uniquely defined by the formulas (\ref{coplevonetxt}) 
(see the text) together with the original ones for the Lie algebra,
admit a triangular decomposition determined by (see comments in the main text and the appendix)
\begin{align}
\label{decozero}
&\Delta(\xi^+_{i,1}) =  \xi^+_{i,1} \otimes 1 + \fU^{[i]} \otimes \xi^+_{i,1} + \xi^+_{i,0} \otimes \kappa_{i,0} + \alg{E} \fU^{[Y]}\otimes Y,\nonumber\\
&\Delta(\xi^-_{i,1}) = \xi^-_{i,1} \otimes 1 + \fU^{-[i]} \otimes \xi^-_{i,1} + \kappa_{i,0} \fU^{-[i]} \otimes \xi^-_{i,0} + Y \otimes \alg{F},\nonumber\\
&\Delta(\kappa_{i,1}) = \kappa_{i,1} \otimes 1 + 1 \otimes \kappa_{i,1} + \kappa_{i,0} \otimes \kappa_{i,0} + \alg{E} \fU^{[\alg{F}]}  \otimes \alg{F}.
\end{align}

The corresponding counits and antipodes, completing the Hopf algebra structure,
are easily derived from the formulas for the coproducts, using the Hopf 
algebra definitions, and they all satisfy the charge conjugation rule 
\cite{J,PST,Gleb}
(see also comments in the appendix).
The realization in terms of Drinfeld currents is given by formulas 
(\ref{curr1txt}), (\ref{curr2txt}) in the text.

We have found the isomorphism between this realization and the one originally given by Beisert in \cite{BYang}. The map can almost be directly guessed from the standard one \cite{Dsecond}, but it nevertheless requires some important modifications, which we comment upon in the main text (see Section \ref{ssec:d2}).
In the Appendix, we constructively rederive the same result starting from a more general transformation on the representation of \cite{BYang}, and imposing suitable constraints.

\section{Yangian $\Ya{\alg{g}}$}\label{sec:yang}

In this section we summarize the definitions for the Yangian $\Ya{\alg{g}}$ of a simple Lie algebra $\alg{g}$ in Drinfeld's first and second realization and give the isomorphism between the two \cite{Drin,Dsecond}. The generalisation to simple Lie superalgebras is straightforward (see for instance \cite{stuko}), however, for simplicity, we refrain from giving the needed modifications in this section. The reader is referred to the standard literature (see for example 
\cite{Etingof,Molev}) for a thorough treatment of this subject.

\subsection{Drinfeld's first realization of $\Ya{\alg{g}}$}\label{ssec:drinf1}

Let $\alg{g}$ be a finite dimensional simple Lie algebra generated by $\gen{J}^A$ satisfying commutation relations $\comm{\gen{J}^A}{\gen{J}^B} = f^{AB}_C\gen{J}^C$, and equipped with a non-degenerate invariant bilinear form $\kappa^{AB}$. The Yangian $\Ya{g}$ is a deformation of the algebra $\alg{g}[u]$ of polynomials with values in $\alg{g}$ defined by the following commutation relations between the level zero generators $\gen{J}^a$ forming $\alg{g}$ and the level one generators $\genY{J}^a$:

\<
\comm{\gen{J}^A}{\gen{J}^B} = f^{AB}_C\gen{J}^C,\\
\comm{\gen{J}^A}{\genY{J}^B} = f^{AB}_C\genY{J}^C.
\>

The generators of higher levels are derived by demanding compatibility with the Serre relation (for algebras strictly larger than $\alg{su(2)}$)

\<
\comm{\genY{J}^{A}}{\comm{\genY{J}^B}{\gen{J}^{C}}} + 
\comm{\genY{J}^{B}}{\comm{\genY{J}^C}{\gen{J}^{A}}} +
\comm{\genY{J}^{C}}{\comm{\genY{J}^A}{\gen{J}^{B}}}
\eq \frac{1}{4}
f^{AG}_{D}f^{BH}_Ef^{CK}_{F}f_{GHK}
\gen{J}^{\{D}\gen{J}^E\gen{J}^{F\}}.
\>

Indices are raised or lowered with the Killing form $\kappa^{AB}$.
The Yangian is equipped with a Hopf algebra structure. The coproduct is uniquely determined for all generators by specifying it on the level zero and one generators as follows

\<
\copro\gen{J}^A\eq\gen{J}^A\otimes 1+1\otimes\gen{J}^A,
\nln
\copro\genY{J}^A\eq\genY{J}^A\otimes 1+1\otimes\genY{J}^A+\half \struc^{A}_{BC}\gen{J}^B\otimes\gen{J}^C.
\>
The antipode and the counit are easily obtained directly from the Hopf algebra definitions.

\subsection{Drinfeld's second realization of $\Ya{\alg{g}}$}\label{ssec:drinf2}

As mentioned in the Introduction, establishing the Yangian $\Ya{\alg{g}}$ in Drinfeld's first realization is not sufficient for the construction of the universal R-matrix, and as well not very suitable for the study of finite dimensional representations. 
Drinfeld's second realization of $\Ya{\alg{g}}$ is defined by generators $\kappa_{i,m}, \xi^\pm_{i,m}$, $i=1,\dots, \text{rank} \alg{g}$, $m=0,1,2,\dots$, with the commutation relations
\begin{align}
\label{def:drinf2}
&[\kappa_{i,m},\kappa_{j,n}]=0,\quad [\kappa_{i,0},\xi^+_{j,m}]=a_{ij} \,\xi^+_{j,m},\nonumber\\
&[\kappa_{i,0},\xi^-_{j,m}]=- a_{ij} \,\xi^-_{j,m},\quad \comm{\xi^+_{j,m}}{\xi^-_{j,n}}=\delta_{i,j}\, \kappa_{j,n+m},\nonumber\\
&[\kappa_{i,m+1},\xi^+_{j,n}]-[\kappa_{i,m},\xi^+_{j,n+1}] = \frac{1}{2} a_{ij} \{\kappa_{i,m},\xi^+_{j,n}\},\nonumber\\
&[\kappa_{i,m+1},\xi^-_{j,n}]-[\kappa_{i,m},\xi^-_{j,n+1}] = - \frac{1}{2} a_{ij} \{\kappa_{i,m},\xi^-_{j,n}\},\nonumber\\
&\comm{\xi^\pm_{i,m+1}}{\xi^\pm_{j,n}}-\comm{\xi^\pm_{i,m}}{\xi^\pm_{j,n+1}} = \pm\frac{1}{2} a_{ij} \acomm{\xi^\pm_{i,m}}{\xi^\pm_{j,n}},\nn\\
&i\neq j,\, \, \, \, n_{ij}=1+|a_{ij}|,\, \, \, \, \, Sym_{\{k\}} [\xi^\pm_{i,k_1},[\xi^\pm_{i,k_2},\dots [\xi^\pm_{i,k_{n_{ij}}}, \xi^\pm_{j,l}]\dots]]=0.
\end{align}
In these formulas $a_{ij}$ is the Cartan matrix, which we will assume to be symmetric throughout the paper.

Having established two a priori independent realizations of the same algebraic structure, one needs to show that the two are isomorphic. Let $\gen{H}_i, \gen{E}_i^\pm$ be a Chevalley-Serre basis for $\alg{g}$, and denote by $\gen{\hat H}_i, \gen{\hat E}_i^\pm$ the appropriate level one generators in the first realization of the Yangian. Then, Drinfeld \cite{Dsecond} gave the isomorphism
\begin{align}
\label{def:isom}
&\kappa_{i,0}=\gen{H}_i,\quad \xi^+_{i,0}=\gen{E}^+_i,\quad \xi^-_{i,0}=\gen{E}^-_i,\nonumber\\
&\kappa_{i,1}=\hat{\gen{H}}_i-v_i,\quad \xi^+_{i,1}=\hat{\gen{E}}^+_i-w_i,\quad \xi^-_{i,1}=\hat{\gen{E}}^-_i-z_i,
\end{align}
where 

\<\label{def:specialel}
v_i &=& \frac{1}{4} \sum_{\beta\in\Delta^+}\left(\alpha_i,\beta\right)(\cwgen{E}_\beta^-\cwgen{E}_\beta^+ +  \cwgen{E}_\beta^+\cwgen{E}_\beta^-) - \half\cwgen{H}_i^2, \\
w_i &=& \frac{1}{4}\sum_{\beta\in\Delta^+}  \left(\cwgen{E}_\beta^-\adjb{\gen{E}_i^+}{\cwgen{E}_\beta^+} + \adjb{\gen{E}_i^+}{\cwgen{E}_\beta^+}\cwgen{E}_\beta^- \right) -  \frac{1}{4}\acomm{\gen{E}_i^+}{\gen{H}_i}, \\
z_i &=& \frac{1}{4}\sum_{\beta\in\Delta^+}  \left(\adjb{\cwgen{E}_\beta^-}{\gen{E}_i^-}\cwgen{E}_\beta^+ + \cwgen{E}_\beta^+ \adjb{\cwgen{E}_\beta^-}{\gen{E}_i^-} \right) -  \frac{1}{4}\acomm{\gen{E}_i^-}{\gen{H}_i} .
\>

Here $\Delta^+$ denotes the set of positive root vectors, $\cwgen{E}^\pm_\beta$ the corresponding generators of the Cartan-Weyl basis, and the adjoint action is defined as $\adjb{x}{y} = [x,y]$. The reader will be able to find literature on the connection between the two realizations for the (related) case of quantum affine algebras, for instance in \cite{unodue}.

\section{The centrally extended Lie superalgebra $\psucentral$}\label{ssec:sucentral}

In this section we will give the definition of the Lie superalgebra $\psucentral$, for a choice of both a Cartan-Weyl and a Chevalley-Serre basis, as the former is suitable for the construction of the Yangian in Drinfeld's first realization, whereas the latter is used to construct the second realization.

\subsection{Chevalley-Serre basis}\label{ssec:CSbasis}

The Lie superalgebra $\psucentral$ is defined in the fermionic Chevalley-Serre basis $\gen{H}_i, \gen{E}^\pm_i$, $i = 1,2,3$, by the standard commutation relations
\<
\comm{\gen{H}_i}{\gen{H}_j} = 0, \\
\comm{\gen{H}_i}{\gen{E}^\pm_j} = \pm {a_{ij}}\gen{E}_j^\pm,  \\
\acomm{\gen{E}^+_i}{\gen{E}^-_j} = \delta_{ij}{\gen{H}_i},
\>
and the Serre relations
\<
\adj{{\gen{E}}_1^\pm}{2}{{\gen{E}}_2^\pm} = \adj{{\gen{E}}_2^\pm}{2}{{\gen{E}}_1^\pm} = 0, \\
\acomm{{\gen{E}}_2^\pm}{\gen{E}_3^\pm} = \text{central} .
\>
The Cartan matrix is given by 

\[
 a_{ij} = 
\begin{pmatrix}
0&-1&1\\
-1&0&0\\
1&0&0
\end{pmatrix}, 
\]
and all roots $\gen{E}_i^\pm$ are fermionic. Furthermore, the curly brackets $\acomm{}{}$ denote the anticommutator, whereas $\adjbb{}$ denotes the super adjoint $\adjb{\gen{J}_i}{\gen{J}_j} = \gen{J}_i\gen{J}_j - (-1)^{|i||j|}\gen{J}_j\gen{J}_i$. Note that the Cartan matrix is degenerate, its null vector $\gen{H}_2 + \gen{H}_3$ being also central. If one sets all three linearly independent central elements to zero, one obtains the simple Lie superalgebra $\psu$.

\subsection{Cartan-Weyl basis}\label{ssec:CWbasis}

Let $\alpha_i$ be the root vectors corresponding to the positive simple roots $\gen{E}^+_i$. Then all positive root vectors are given by $\beta_1 = \alpha_2, \beta_2 = \alpha_1 + \alpha_2, \beta_3 = \alpha_1 + \alpha_2 + \alpha_3, \beta_4 = \alpha_1, \beta_5 = \alpha_1 + \alpha_3, \beta_6 = \alpha_3$.
The root vector $\beta_7 = \alpha_2 + \alpha_3$ has zero length, but nevertheless corresponds to some generator, namely a central element. 
The corresponding Cartan Weyl basis reads
\begin{align}
\cwgen{E}^+_{\beta_1} &= \gen{E}^+_2,\qquad &\cwgen{E}^-_{\beta_1} &= \gen{E}^-_2 \\
\cwgen{E}^+_{\beta_2} &= \acomm{\gen{E}^+_1}{\gen{E}^+_2},\qquad &\cwgen{E}^-_{\beta_2} &= \acomm{\gen{E}^-_1}{\gen{E}^-_2} \\
\cwgen{E}^+_{\beta_3} &= \comm{\acomm{\gen{E}^+_1}{\gen{E}^+_2}}{\gen{E}^+_3},\qquad &\cwgen{E}^-_{\beta_3} &= \comm{\acomm{\gen{E}^-_1}{\gen{E}^-_2}}{\gen{E}^-_3}, \\
\cwgen{E}^+_{\beta_4} &= \gen{E}^+_1,\qquad &\cwgen{E}^-_{\beta_4} &= \gen{E}^-_1 \\
\cwgen{E}^+_{\beta_5} &=  \acomm{\gen{E}^+_1}{\gen{E}^+_3},\qquad &\cwgen{E}^-_{\beta_5} &=  \acomm{\gen{E}^-_1}{\gen{E}^-_3} \\
\cwgen{E}^+_{\beta_6} &= \gen{E}^+_3 \qquad &\cwgen{E}^-_{\beta_6} &= \gen{E}^-_3,\\
\cwgen{E}^+_{\beta_7} &=  \acomm{\gen{E}^+_2}{\gen{E}^+_3},\qquad &\cwgen{E}^-_{\beta_7} &=  \acomm{\gen{E}^-_2}{\gen{E}^-_3}.
\end{align}

For later purposes, we define $a_i$ such that $\comm{\cwgen{E}_i^+}{\cwgen{E}_i^-} = a_i\cwgen{H}_i$ holds. $\cwgen{H}_i$ is the dual to $\beta_i$, namely, if $\beta_i = \alpha_k + \alpha_l + \dots$, then $\cwgen{H}_{\beta_i} = \gen{H}_k + \gen{H}_l + \dots$. In this way $a_i$ is given by 

\[\label{def:cwa}
a_i = \{1,1,1,1,-1,1\}.
\]

The definition given by Beisert \cite{B,BYang} of the centrally extended 
$\alg{su(2|2)}$ Lie superalgebra is as follows. It is generated by two $\alg{su(2)}$'s denoted by $\gen{R}^{a}{}_{b}$,
$\gen{L}^{\alpha}{}_{\beta}$, eight supercharges
$\gen{Q}^{\alpha}{}_{b}$, $\gen{S}^{a}{}_{\beta}$ and three central
charges $\gen{C}$, $\gen{P}$, $\gen{K}$ satisfying the following (non-vanishing) commutation relations:
\[
\begin{array}[b]{rclcrcl}
\comm{\gen{R}^a{}_b}{\gen{R}^c{}_d}\eq
\delta^c_b\gen{R}^a{}_d-\delta^a_d\gen{R}^c{}_b,
&&
\comm{\gen{L}^\alpha{}_\beta}{\gen{L}^\gamma{}_\delta}\eq
\delta^\gamma_\beta\gen{L}^\alpha{}_\delta-\delta^\alpha_\delta\gen{L}^\gamma{}_\beta,
\\[3pt]
\comm{\gen{R}^a{}_b}{\gen{Q}^\gamma{}_d}\eq
-\delta^a_d\gen{Q}^\gamma{}_b+\half \delta^a_b\gen{Q}^\gamma{}_d,
&&
\comm{\gen{L}^\alpha{}_\beta}{\gen{Q}^\gamma{}_d}\eq
+\delta^\gamma_\beta\gen{Q}^\alpha{}_d-\half \delta^\alpha_\beta\gen{Q}^\gamma{}_d,
\\[3pt]
\comm{\gen{R}^a{}_b}{\gen{S}^c{}_\delta}\eq
+\delta^c_b\gen{S}^a{}_\delta-\half \delta^a_b\gen{S}^c{}_\delta,
&&
\comm{\gen{L}^\alpha{}_\beta}{\gen{S}^c{}_\delta}\eq
-\delta^\alpha_\delta\gen{S}^c{}_\beta+\half \delta^\alpha_\beta\gen{S}^c{}_\delta,\nonumber
\end{array}\]
\<
\acomm{\gen{Q}^\alpha{}_b}{\gen{S}^c{}_\delta}\eq
\delta^c_b\gen{L}^\alpha{}_\delta +\delta^\alpha_\delta\gen{R}^c{}_b
+\delta^c_b\delta^\alpha_\delta\gen{C},
\nln
\acomm{\gen{Q}^{\alpha}{}_{b}}{\gen{Q}^{\gamma}{}_{d}}\eq
\varepsilon^{\alpha\gamma}\varepsilon_{bd}\gen{P},
\nln
\acomm{\gen{S}^{a}{}_{\beta}}{\gen{S}^{c}{}_{\delta}}\eq
\varepsilon^{ac}\varepsilon_{\beta\delta}\gen{K}.
\>
All indices run from $1$ to $2$ and we have the additional trace condition $\gen{R}^1_1 + \gen{R}^2_2 = \gen{L}^1_1 + \gen{L}^2_2 = 0$. 
This definition of $\psucentral$ coincides with the Chevalley-Serre 
presentation if one sets
\begin{align}
\gen{E}_1^+ &= \gen{Q}^2_2, \qquad &\gen{E}_1^- &= \gen{S}^2_2, \qquad &\gen{H}_1 & = -\gen{R}_1^1 - \gen{L}^1_1 + \gen{C},\\
\gen{E}_2^+ &= i\gen{S}^1_2, \qquad &\gen{E}_2^- &= i\gen{Q}^2_1, \qquad &\gen{H}_2 & = -\gen{R}_1^1 + \gen{L}^1_1 - \gen{C},\\
\gen{E}_3^+ &= i\gen{S}^2_1, \qquad &\gen{E}_3^- &= i\gen{Q}^1_2, \qquad &\gen{H}_3 & = \gen{R}_1^1 - \gen{L}^1_1 - \gen{C} , 
\end{align}
and with the Cartan-Weyl description by setting
\begin{align}
\cwgen{E}_1^+ &= i\gen{S}^1_2, \qquad &\cwgen{E}_1^- &= i\gen{Q}^2_1, \qquad &\cwgen{H}_1 & = -\gen{R}_1^1 + \gen{L}^1_1 - \gen{C},\\
\cwgen{E}_2^+ &= i\gen{R}^1_2, \qquad &\cwgen{E}_2^- &= i\gen{R}^2_1, \qquad &\cwgen{H}_2 & = -\gen{R}_1^1 + \gen{R}^2_2,\\
\cwgen{E}_3^+ &= -\gen{S}^1_1, \qquad &\cwgen{E}_3^- &= \gen{Q}^1_1, \qquad &\cwgen{H}_3 & = -\gen{R}_1^1 - \gen{L}^1_1 - \gen{C},\\
\cwgen{E}_4^+ &= \gen{Q}^2_2, \qquad &\cwgen{E}_4^- &= \gen{S}^2_2, \qquad &\cwgen{H}_4 & = -\gen{R}_1^1 - \gen{L}^1_1 + \gen{C},\\
\cwgen{E}_5^+ &= i\gen{L}^2_1, \qquad &\cwgen{E}_5^- &= i\gen{L}^1_2, \qquad &\cwgen{H}_5 & = -\gen{L}_1^1 + \gen{L}^2_2,\\
\cwgen{E}_6^+ &= i\gen{S}^2_1, \qquad &\cwgen{E}_6^- &= i\gen{Q}^1_2, \qquad &\cwgen{H}_6 & = \gen{R}_1^1 - \gen{L}^1_1 - \gen{C} .
\end{align}

The additional central elements are given by 

\[
 \gen{K} = \cwgen{E}^+_7 = \acomm{\gen{E}^+_3}{\gen{E}^+_2} ,\quad \gen{P} = \cwgen{E}^-_7 = \acomm{\gen{E}^-_2}{\gen{E}^-_3} .
\]

The fundamental four-dimensional representation of $\psucentral$ is defined by \cite{B}
\begin{align}\label{def:funrep}
&\fR^a{}_b|\phi^c\rangle=\delta^c_b|\phi^a\rangle - \frac{1}{2} \delta^a_b|\phi^c\rangle,\quad 
\fL^\alpha{}_\beta|\phi^\gamma\rangle=\delta^\gamma_\beta|\phi^\alpha\rangle - \frac{1}{2} \delta^\alpha_\beta|\phi^\gamma\rangle,\nonumber\\ 
&\fQ^\alpha{}_a|\phi^b\rangle=a\, \delta^b_a|\psi^\alpha\rangle,\quad
\fQ^\alpha{}_a|\psi^\beta\rangle
=b\, \epsilon^{\alpha\beta}\epsilon_{ab}|\phi^b\rangle,\nonumber\\
&\fS^a{}_\alpha|\phi^b\rangle
=c\, \epsilon^{ab}\epsilon_{\alpha\beta}|\psi^\beta\rangle,\quad
\fS^a{}_\alpha|\psi^\beta\rangle=d\, \delta^\beta_\alpha|\phi^a\rangle,
\end{align}
where the four-dimensional vector space is spanned by two bosons $|\phi^a\rangle, a=1,2$ and two fermions $|\psi^\alpha\rangle, \alpha=1,2$. The four complex numbers $a,b,c,d$ labeling the representation have to satisfy the constraint $ad-bc=1$. We solve this constraint as $d=\frac{1 + bc}{a}$ for definiteness in what follows. The eigenvalues of the central charges are given by 
$C=\frac{1}{2}+bc$, $P=ab$ and $K=cd$. For more on the representation theory of $\psucentral$ see \cite{nlin}.

\section{The Yangian $\Ya{\psucentral}$}

In this section we will first review Drinfeld's first, and then derive Drinfeld's second realization for the Yangian $\Ya{\psucentral}$. This can be done in a similar way as for ordinary simple Lie algebras, as outlined in Section \ref{sec:yang}. However, there are some crucial differences which we will point out.

\subsection{Drinfeld's first realization of $\Ya{\psucentral}$}

In Section \ref{ssec:drinf1} we reviewed how to construct the Yangian $\Ya{\alg{g}}$ based on a simple Lie algebra. One faces the problem that $\psucentral$ is not simple. The procedure by Beisert \cite{BYang} was based on the following arguments. Essential to the construction is a non-degenerate invariant bilinear form. The Killing form for $\psucentral$ being degenerate, the $\alg{sl(2)}$ automorphisms were adjoined to the algebra, yielding the algebra $\alg{sl(2)}\ltimes\psucentral$, which possesses a non-degenerate form. As these automorphisms drop out of the coproduct for generators of $\psucentral$, this produced a formal definition of the Yangian $\Ya{\psucentral}$ without explicitely mentioning the automorphisms. However, they are implicitely used to lower the indices of the structure constants. We spell out the results obtained by Beisert in $\cite{BYang}$ for the case where the coproduct is twisted by an additional braiding element $\gen{U}$, which was first derived in \cite{GH,PST} for the level zero generators. The coproduct has then the form 
\<\label{def:drinf1copro}
\copro\gen{J}^A &=&
\gen{J}^A\otimes 1+\eip^{[A]}\otimes\gen{J}^A,\\
\copro\genY{J}^A &=&
\genY{J}^A\otimes 1+\eip^{[A]}\otimes\genY{J}^A
+\struc^{A}_{BC}\gen{J}^B\eip^{[C]}\otimes\gen{J}^C ,
\>
where $[A]$ denotes the ``braid''-charge, which is the eigenvalue of any generator with respect to the Cartan generator $\gen{B}^1_1 - \gen{B}^2_2$ of the $\alg{sl(2)}$ automorphism, namely

\[
 \comm{\gen{B}^1_1 - \gen{B}^2_2}{\gen{J}^A} = [A]\gen{J}^A .
\]
Beisert's result is reported explicitely in Table \ref{tab:drinf1co}.

\begin{table}\centering
\<
\copro\genY{R}^a{}_b\eq
\genY{R}^a{}_b\otimes 1
+1\otimes\genY{R}^a{}_b
\nl
+\half\gen{R}^a{}_c\otimes\gen{R}^c{}_b
-\half\gen{R}^c{}_b\otimes\gen{R}^a{}_c
\nl
-\half\gen{S}^a{}_\gamma\eip^{+1}\otimes\gen{Q}^\gamma{}_b
-\half\gen{Q}^\gamma{}_b\eip^{-1}\otimes\gen{S}^a{}_\gamma
\nl\qquad
+\quarter\delta^a_b\,\gen{S}^d{}_\gamma\eip^{+1}\otimes\gen{Q}^\gamma{}_d
+\quarter\delta^a_b\,\gen{Q}^\gamma{}_d\eip^{-1}\otimes\gen{S}^d{}_\gamma,
\nln
\copro\genY{L}^\alpha{}_\beta\eq
\genY{L}^\alpha{}_\beta\otimes 1
+1\otimes\genY{L}^\alpha{}_\beta
\nl
-\half\gen{L}^\alpha{}_\gamma\otimes\gen{L}^\gamma{}_\beta
+\half\gen{L}^\gamma{}_\beta\otimes\gen{L}^\alpha{}_\gamma
\nl
+\half\gen{Q}^\alpha{}_c\eip^{-1}\otimes\gen{S}^c{}_\beta
+\half\gen{S}^c{}_\beta\eip^{+1}\otimes\gen{Q}^\alpha{}_c
\nl\qquad
-\quarter\delta^\alpha_\beta\,\gen{Q}^\delta{}_c\eip^{-1}\otimes\gen{S}^c{}_\delta
-\quarter\delta^\alpha_\beta\,\gen{S}^c{}_\delta\eip^{+1}\otimes\gen{Q}^\delta{}_c,
\nln
\copro\genY{Q}^\alpha{}_b\eq
\genY{Q}^\alpha{}_b\otimes 1
+\eip^{+1}\otimes\genY{Q}^\alpha{}_b
\nl
-\half\gen{L}^\alpha{}_\gamma\eip^{+1}\otimes\gen{Q}^\gamma{}_b
+\half\gen{Q}^\gamma{}_b\otimes\gen{L}^\alpha{}_\gamma
\nl
-\half\gen{R}^c{}_b\eip^{+1}\otimes\gen{Q}^\alpha{}_c
+\half\gen{Q}^\alpha{}_c\otimes\gen{R}^c{}_b
\nl
-\half\gen{C}\eip^{+1}\otimes\gen{Q}^\alpha{}_b
+\half\gen{Q}^\alpha{}_b\otimes\gen{C}
\nl
+\half\varepsilon^{\alpha\gamma}\varepsilon_{bd}\gen{P}\eip^{-1}\otimes\gen{S}^d{}_\gamma
-\half\varepsilon^{\alpha\gamma}\varepsilon_{bd}\gen{S}^d{}_\gamma\eip^{+2}\otimes\gen{P},
\nln
\copro\genY{S}^a{}_\beta\eq
\genY{S}^a{}_\beta\otimes 1
+\eip^{-1}\otimes\genY{S}^a{}_\beta
\nl
+\half\gen{R}^a{}_c\eip^{-1}\otimes\gen{S}^c{}_\beta
-\half\gen{S}^c{}_\beta\otimes\gen{R}^a{}_c
\nl
+\half\gen{L}^\gamma{}_\beta\eip^{-1}\otimes\gen{S}^a{}_\gamma
-\half\gen{S}^a{}_\gamma\otimes\gen{L}^\gamma{}_\beta
\nl
+\half\gen{C}\eip^{-1}\otimes\gen{S}^a{}_\beta
-\half\gen{S}^a{}_\beta\otimes\gen{C}
\nl
-\half\varepsilon^{ac}\varepsilon_{\beta\delta}\gen{K}\eip^{+1}\otimes\gen{Q}^\delta{}_c
+\half\varepsilon^{ac}\varepsilon_{\beta\delta}\gen{Q}^\delta{}_c\eip^{-2}\otimes\gen{K},
\nln
\copro\genY{C}\eq
\genY{C}\otimes 1
+1\otimes\genY{C}
\nl
+\half(\gen{P}\eip^{-2}\otimes\gen{K}
-\gen{K}\eip^{+2}\otimes\gen{P}) ,
\nln
\copro\genY{P}\eq
\genY{P}\otimes 1
+\eip^{+2}\otimes\genY{P}
\nl
-\gen{C}\eip^{+2}\otimes\gen{P}
+\gen{P}\otimes\gen{C},
\nln
\copro\genY{K}\eq
\genY{K}\otimes 1
+\eip^{-2}\otimes\genY{K}
\nl
+\gen{C}\eip^{-2}\otimes\gen{K}
-\gen{K}\otimes\gen{C}.
\nn
\>
\caption{The coproduct for $\grp{Y}(\alg{h})$ in Drinfeld's first realization.}
\label{tab:drinf1co}
\end{table}

\subsection{Drinfeld's second realization of $\Ya{\psucentral}$}\label{ssec:d2}

We define $\Ya{\psucentral}$ in the second realization in almost the same way as it is done for an arbitrary simple Lie algebra, as outlined in Section \ref{ssec:drinf2}. It is defined in terms of generators $\kappa_{i,m}, \xi^\pm_{i,m}$, $i=1,2,3, m=0,1,2,\dots$, with the commutation relations

\begin{align}
\label{def:drinf2su22}
&[\kappa_{i,m},\kappa_{j,n}]=0,\quad [\kappa_{i,0},\xi^+_{j,m}]=a_{ij} \,\xi^+_{j,m},\nonumber\\
&[\kappa_{i,0},\xi^-_{j,m}]=- a_{ij} \,\xi^-_{j,m},\quad \{\xi^+_{j,m},\xi^-_{j,n}\}=\delta_{i,j}\, \kappa_{j,n+m},\nonumber\\
&[\kappa_{i,m+1},\xi^+_{j,n}]-[\kappa_{i,m},\xi^+_{j,n+1}] = \frac{1}{2} a_{ij} \{\kappa_{i,m},\xi^+_{j,n}\},\nonumber\\
&[\kappa_{i,m+1},\xi^-_{j,n}]-[\kappa_{i,m},\xi^-_{j,n+1}] = - \frac{1}{2} a_{ij} \{\kappa_{i,m},\xi^-_{j,n}\},\nonumber\\
&\{\xi^+_{i,m+1},\xi^+_{j,n}\}-\{\xi^+_{i,m},\xi^+_{j,n+1}\} = \frac{1}{2} a_{ij} [\xi^+_{i,m},\xi^+_{j,n}],\nonumber\\
&\{\xi^-_{i,m+1},\xi^-_{j,n}\}-\{\xi^-_{i,m},\xi^-_{j,n+1}\} = - \frac{1}{2} a_{ij} [\xi^-_{i,m},\xi^-_{j,n}],
\end{align}
and the Serre relations 
\begin{eqnarray}
\label{sere}
&&i\neq j, \, \, \, \, \, n_{ij}=1+|a_{ij}|,\, \, \, \, \, Sym_{\{k\}} [\xi^+_{i,k_1},[\xi^+_{i,k_2},\dots \{\xi^+_{i,k_{n_{ij}}}, \xi^+_{j,l}\}\dots\}\}=0,\nonumber\\
&&i\neq j, \, \, \, \, \, n_{ij}=1+|a_{ij}|,\, \, \, \, \, Sym_{\{k\}} [\xi^-_{i,k_1},[\xi^-_{i,k_2},\dots \{\xi^-_{i,k_{n_{ij}}}, \xi^-_{j,l}\}\dots\}\}=0,\nonumber\\
&&\text{except for} \, \, \, \, \, \, \, \, \, \{\xi^+_{2,n},\xi^+_{3,m}\} = \alg{K}_{n+m}, \qquad  \{\xi^-_{2,n},\xi^-_{3,m}\} = \alg{P}_{n+m}, 
\end{eqnarray}
where $\alg{K}_{n}$ and $\alg{P}_{n}$ are central elements, and $a_{ij}$ is the Cartan matrix of $\psucentral$ as given in Section \ref{ssec:CSbasis}. Further, $[x,y\} = xy - (-1)^{|x||y|} yx$ is the supercommutator.

Indeed, the last Serre relation is the only difference with respect to the standard definition of Drinfeld's second realization for Yangians of simple Lie algebras. This realization is close to the Chevalley-Serre presentation of the underlying Lie algebra, where we have seen (cfr. Section \ref{ssec:CSbasis}) that the crucial difference to the simple case is the modification of the Serre relations. In view of this fact, our definition of the second realization of $\Ya{\psucentral}$ is naturally expected. 

We now want to establish the isomorphism between the two realizations of $\Ya{\psucentral}$. What we find is that the structure of the isomorphism is the same as in the case of simple Lie algebras, namely it is given by 
\begin{align}
\label{mappa1}
&\kappa_{i,0}=\gen{H}_i,\quad \xi^+_{i,0}=\gen{E}_i,\quad \xi^-_{i,0}=\gen{F}_i,\nonumber\\
&\kappa_{i,1}=\hat{\gen{H}}_i-v_i,\quad \xi^+_{i,1}=\hat{\gen{E}}_i-w_i,\quad \xi^-_{i,1}=\hat{\gen{F}}_i-z_i,
\end{align}
where the special elements are given by

\begin{eqnarray}
v_1 &=& - \frac{1}{2} \kappa_{1,0}^2 +  \frac{1}{4} (\fR^2{}_1 \fR^1{}_2 +  \fR^1{}_2 \fR^2{}_1 +  \fL^2{}_1 \fL^1{}_2 +  \fL^1{}_2 \fL^2{}_1 \nonumber\\
&&+ \fQ^2{}_1 \fS^1{}_2 - \fQ^1{}_2 \fS^2{}_1 -  \fS^1{}_2 \fQ^2{}_1  + \fS^2{}_1 \fQ^1{}_2) + \half\gen{P}\gen{K} ,\nonumber\\
v_2 &=& - \frac{1}{2} \kappa_{2,0}^2 + \frac{1}{4} (\fR^2{}_1 \fR^1{}_2 +  \fR^1{}_2 \fR^2{}_1 -  \fL^2{}_1 \fL^1{}_2 -  \fL^1{}_2 \fL^2{}_1 \nonumber\\
&&+ \fQ^1{}_1 \fS^1{}_1 + \fQ^2{}_2 \fS^2{}_2 -  \fS^1{}_1 \fQ^1{}_1  - \fS^2{}_2 \fQ^2{}_2) -  \half\gen{P}\gen{K}  ,\nonumber\\  
v_3 &=& - \frac{1}{2} \kappa_{3,0}^2 + \frac{1}{4} (- \fR^2{}_1 \fR^1{}_2 -  \fR^1{}_2 \fR^2{}_1 +  \fL^2{}_1 \fL^1{}_2 +  \fL^1{}_2 \fL^2{}_1 \nonumber\\
&&- \fQ^1{}_1 \fS^1{}_1 - \fQ^2{}_2 \fS^2{}_2 +  \fS^1{}_1 \fQ^1{}_1  + \fS^2{}_2 \fQ^2{}_2) -  \half\gen{P}\gen{K},\nonumber  
\end{eqnarray}
\begin{eqnarray}
\label{suspecial}
w_1 &=& - \frac{1}{4} (\xi^+_{1,0} \kappa_{1,0} + \kappa_{1,0} \xi^+_{1,0}) + \frac{1}{4} (\fQ^2{}_1 \fR^1{}_2 +  \fR^1{}_2 \fQ^2{}_1 +  \fQ^1{}_2 \fL^2{}_1 +  \fL^2{}_1 \fQ^1{}_2 + 2 \fS^1{}_1 \fP) +  \half\gen{S}_1^1\gen{P},\nonumber\\
w_2 &=& - \frac{1}{4} (\xi^+_{2,0} \kappa_{2,0} + \kappa_{2,0} \xi^+_{2,0}) + \frac{i}{4} (\fS^1{}_1 \fL^1{}_2 +  \fL^1{}_2 \fS^1{}_1 -  \fS^2{}_2 \fR^1{}_2 -  \fR^1{}_2 \fS^2{}_2 - 2 \fQ^1{}_2 \fK),\nonumber\\
w_3 &=& - \frac{1}{4} (\xi^+_{3,0} \kappa_{3,0} + \kappa_{3,0} \xi^+_{3,0}) + \frac{i}{4} (\fS^1{}_1 \fR^2{}_1 +  \fR^2{}_1 \fS^1{}_1 -  \fS^2{}_2 \fL^2{}_1 -  \fL^2{}_1 \fS^2{}_2 - 2 \fQ^2{}_1 \fK),\nonumber\\
z_1 &=& - \frac{1}{4} (\xi^-_{1,0} \kappa_{1,0} + \kappa_{1,0} \xi^-_{1,0}) + \frac{1}{4} (\fS^1{}_2 \fR^2{}_1 +  \fR^2{}_1 \fS^1{}_2 + \fS^2{}_1 \fL^1{}_2 +  \fL^1{}_2 \fS^2{}_1 + 2 \fQ^1{}_1 \fK)+ \half\gen{Q}_1^1\gen{K},\nonumber\\
z_2 &=& - \frac{1}{4} (\xi^-_{2,0} \kappa_{2,0} + \kappa_{2,0} \xi^-_{2,0}) + \frac{i}{4} (\fQ^1{}_1 \fL^2{}_1 +  \fL^2{}_1 \fQ^1{}_1 -  \fQ^2{}_2 \fR^2{}_1 -  \fR^2{}_1 \fQ^2{}_2 - 2 \fS^2{}_1 \fP),\nonumber\\
z_3 &=& - \frac{1}{4} (\xi^-_{3,0} \kappa_{3,0} + \kappa_{3,0} \xi^-_{3,0}) + \frac{i}{4} (\fQ^1{}_1 \fR^1{}_2 +  \fR^1{}_2 \fQ^1{}_1 -  \fQ^2{}_2 \fL^1{}_2 -  \fL^1{}_2 \fQ^2{}_2 - 2 \fS^1{}_2 \fP).\nonumber\\
\end{eqnarray}
However, we note that these special elements do not exactly have the standard form 

\<\label{def:superspecial}
(v_i)_{\text{standard}} &=& \frac{1}{4} \sum_{\beta\in\Delta^+}\left(\alpha_i,\beta\right)(\cwgen{E}_\beta^-\cwgen{E}_\beta^+ +  (-1)^{|\cwgen{E}_\beta^-|}\cwgen{E}_\beta^+\cwgen{E}_\beta^-) - \half\gen{H}_i^2, \nn\\
(w_i)_{\text{standard}} &=& \frac{1}{4}\sum_{\beta\in\Delta^+}a_\beta\left(\cwgen{E}_\beta^-\adjb{\gen{E}_i^+}{\cwgen{E}_\beta^+} + (-1)^{|\beta|(|\beta|+|i|)}\adjb{\gen{E}_i^+}{\cwgen{E}_\beta^+}\cwgen{E}_\beta^- \right) -  \frac{1}{4}\acomm{\gen{E}_i^+}{\gen{H}_i}, \nn\\
(z_i)_{\text{standard}} &=& \frac{1}{4}\sum_{\beta\in\Delta^+}a_\beta(-1)^{|\beta|}\left(\adjb{\cwgen{E}_\beta^-}{\gen{E}_i^-}\cwgen{E}_\beta^+ + (-1)^{|\beta|(|\beta|+|i|)}\cwgen{E}_\beta^+ \adjb{\cwgen{E}_\beta^-}{\gen{E}_i^-} \right) -  \frac{1}{4}\acomm{\gen{E}_i^-}{\gen{H}_i},\nn\\
\>
where we have promoted the formulas 
given in Section \ref{ssec:drinf2} for the isomorphism between the two 
realizations to the case of superalgebras, and used the elements $a_\beta$ as defined in (\ref{def:cwa}).

The non-zero differences are given by the following terms: 

\<
(v_1)_{\text{standard}} - v_1 &=& - \half\gen{P}\gen{K},\\
(v_2)_{\text{standard}} - v_2 &=& + \half\gen{P}\gen{K},\\
(v_3)_{\text{standard}} - v_3 &=& + \half\gen{P}\gen{K},\\
(w_1)_{\text{standard}} - w_1 &=& - \half\gen{S}_1^1\gen{P},\\
(z_1)_{\text{standard}} - z_1 &=& - \half\gen{Q}_1^1\gen{K}.
\>

Let us comment on the difference between our result and the standard case. As $\gen{K}$ corresponds to the root vector $\alpha_3 + \alpha_2$, which has zero scalar product with all other root vectors, the extra terms $\pm\half\gen{P}\gen{K}$ in $v_i$ cannot be obtained directly from the general formulas (\ref{def:superspecial}). It is interesting to notice that, in $\alg{d}(2,1;\alpha)$, the scalar product of the roots which in the $\alpha\rightarrow 0$ limit (see e.g. \cite{B}) become $\gen{P},\gen{K}$ is nonzero. It would be interesting to see if one could get these special elements for $\Ya{\psucentral}$ as a contraction from $\Ya{\alg{d(2,1;\alpha)}}$. For more work on the relation between $\alg{d(2,1;\alpha)}$ and $\psucentral$ see \cite{Moriyama}, and \cite{HSTY}, whose treatment goes along the presentation of \cite{Yamane}. 

For the roots $2$ and $3$ the isomorphisms are precisely the standard ones. The appearance of the shifts for $w_1, z_1$ can be understood as follows. When $\gen{P}$, $\gen{K}$ are not zero, the anticommutator of the positive simple root $\cwgen{E}_4^+$ with the negative root $\cwgen{E}^-_3$ is non-zero and gives the central element $\gen{P}$. Hence we interpret the shift term as $-\half\gen{S}_1^1\gen{P} = \frac{1}{4}(\cwgen{E}_{\beta_3}^+\adjb{\gen{E}_1^+}{\cwgen{E}_{\beta_3}^-} + \adjb{\gen{E}_1^+}{\cwgen{E}_{\beta_3}^-}\cwgen{E}_{\beta_3}^+)$, i.e. we include the negative root $-\beta_3$ in the sum in (\ref{def:superspecial}). This is in principle consistent since the root diagram is only two dimensional, due to the degeneracy of the Cartan matrix. Again, one may also try to make sense of this term when considering the contraction from $\alg{d}(2,1;\alpha)$.

\subsection{Coalgebra structure}
It is important to provide the coproducts for the Yangian generators, 
which respect the defining relations we have given. Such a coalgebra structure 
is uniquely defined by the coproducts on the level zero generators, and the following ones on the level one generators:

\begin{eqnarray}
\label{coplevonetxt}
\Delta(\kappa_{1,1}) &=& \kappa_{1,1} \otimes 1 + 1 \otimes \kappa_{1,1} + \kappa_{1,0} \otimes \kappa_{1,0} - \fK \fU^2 \otimes \fP - \fR^1{}_2 \otimes \fR^2{}_1- \fL^2{}_1 \otimes \fL^1{}_2 \nonumber\\
&&+ \fS^1{}_2 \fU \otimes \fQ^2{}_1 -  \fS^2{}_1 \fU \otimes \fQ^1{}_2,\nonumber\\ 
\Delta(\kappa_{2,1}) &=& \kappa_{2,1} \otimes 1 + 1 \otimes \kappa_{2,1} + \kappa_{2,0} \otimes \kappa_{2,0} + \fK \fU^2 \otimes \fP - \fR^1{}_2 \otimes \fR^2{}_1 + \fL^2{}_1 \otimes \fL^1{}_2 \nonumber\\
&&+ \fS^1{}_1 \fU \otimes \fQ^1{}_1 -  \fQ^2{}_2 \fU^{-1} \otimes \fS^2{}_2,\nonumber\\ 
\Delta(\kappa_{3,1}) &=& \kappa_{3,1} \otimes 1 + 1 \otimes \kappa_{3,1} + \kappa_{3,0} \otimes \kappa_{3,0} + \fK \fU^2 \otimes \fP + \fR^1{}_2 \otimes \fR^2{}_1 - \fL^2{}_1 \otimes \fL^1{}_2 \nonumber\\
&&- \fS^1{}_1 \fU \otimes \fQ^1{}_1 +  \fQ^2{}_2 \fU^{-1} \otimes \fS^2{}_2,\nonumber\\ 
\Delta(\xi^+_{1,1}) &=& \xi^+_{1,1} \otimes 1 + \fU \otimes \xi^+_{1,1} + \xi^+_{1,0} \otimes \kappa_{1,0} - \fS^1{}_1 \fU^2 \otimes \fP - \fR^1{}_2 \fU \otimes \fQ^2{}_1 - \fL^2{}_1 \fU \otimes \fQ^1{}_2, \nonumber\\
\Delta(\xi^+_{2,1}) &=& \xi^+_{2,1} \otimes 1 + \fU^{-1} \otimes \xi^+_{2,1} + \xi^+_{2,0} \otimes \kappa_{2,0} + i \fK \fU \otimes \fQ^1{}_2  - i \fS^1{}_1 \otimes \fL^1{}_2 + i \fR^1{}_2 \fU^{- 1} \otimes \fS^2{}_2, \nonumber\\
\Delta(\xi^+_{3,1}) &=& \xi^+_{3,1} \otimes 1 + \fU^{-1} \otimes \xi^+_{3,1} + \xi^+_{3,0} \otimes \kappa_{3,0} + i \fK \fU \otimes \fQ^2{}_1  - i \fS^1{}_1 \otimes \fR^2{}_1 + i \fL^2{}_1 \fU^{- 1} \otimes \fS^2{}_2, \nonumber\\
\Delta(\xi^-_{1,1}) &=& \xi^-_{1,1} \otimes 1 + \fU^{-1} \otimes \xi^-_{1,1} + \kappa_{1,0} \fU^{- 1} \otimes \xi^-_{1,0} - \fK \fU \otimes \fQ^1{}_1  - \fS^2{}_1 \otimes \fL^1{}_2 - \fS^1{}_2 \otimes \fR^2{}_1, \nonumber\\
\Delta(\xi^-_{2,1}) &=& \xi^-_{2,1} \otimes 1 + \fU \otimes \xi^-_{2,1} + \kappa_{2,0} \fU \otimes \xi^-_{2,0} + i \fS^2{}_1 \fU^2 \otimes \fP  + i \fQ^2{}_2 \otimes \fR^2{}_1 - i \fL^2{}_1 \fU \otimes \fQ^1{}_1, \nonumber\\
\Delta(\xi^-_{3,1}) &=& \xi^-_{3,1} \otimes 1 + \fU \otimes \xi^-_{3,1} + \kappa_{3,0} \fU \otimes \xi^-_{3,0} + i \fS^1{}_2 \fU^2 \otimes \fP  - i \fR^1{}_2 \fU \otimes \fQ^1{}_1 + i \fQ^2{}_2 \otimes \fL^1{}_2. \nonumber\\
\end{eqnarray}
One can notice that these coproducts satisfy a triangular decomposition in the 
spirit of the treatment in 
\cite{KT}, namely they are schematically of the form
\begin{align}
\label{decozero2}
&\Delta(\xi^+_{i,1}) =  \xi^+_{i,1} \otimes 1 + \fU^{[i]} \otimes \xi^+_{i,1} + \xi^+_{i,0} \otimes \kappa_{i,0} + \alg{E} \fU^{[Y]}\otimes Y,\nonumber\\
&\Delta(\xi^-_{i,1}) = \xi^-_{i,1} \otimes 1 + \fU^{-[i]} \otimes \xi^-_{i,1} + \kappa_{i,0} \fU^{-[i]} \otimes \xi^-_{i,0} + Y \otimes \alg{F},\nonumber\\
&\Delta(\kappa_{i,1}) = \kappa_{i,1} \otimes 1 + 1 \otimes \kappa_{i,1} + \kappa_{i,0} \otimes \kappa_{i,0} + \alg{E} \fU^{[\alg{F}]}  \otimes \alg{F},
\end{align}
where $\alg{E}$ and $\alg{F}$ are the subalgebras generated by the positive and negative roots $I_+$ and $I_-$ 
respectively, and $Y$ is the whole Yangian. The notation with 
sets in (\ref{decozero2}) indicates a sum of combinations of terms taken from the 
respective sets. The exponents of the corresponding braiding factors are of obvious 
meaning.

This triangular decomposition is easily checked if one recalls the subdivision
 in positive and negative roots from Section \ref{ssec:sucentral}. 
In particular, the
non-simple roots $\fR^1{}_2$, $\fL^2{}_1$ and $\fS^1{}_1$  
are positive,
while  $\fR^2{}_1$, $\fL^1{}_2$ and $\fQ^1{}_1$ are negative. Furthermore,
$\fK$ is generated inside $I_+$, and $\fP$ inside $I_-$. 

\section{Fundamental Representation}
We are now ready to determine the fundamental evaluation representation of the 
Yangian, in the second realization we have derived. One can convince oneself 
that the following four-dimensional matrix representation
\begin{align}
\label{mult1txt}
&\kappa_{i,n} = \omega_i^n \kappa_{i,0},\quad \xi^+_{i,n} = \omega_i^n \xi^+_{i,0},\quad \xi^-_{i,n} = \omega_i^n \xi^-_{i,0},
\end{align}
with 
\begin{align}
\label{mult2txt}
&\omega_1= ig u,\nonumber\\
&\omega_2 = \omega_3 = ig u - C,
\end{align}
and where $\kappa_{i,0}, \xi^\pm_{i,0}$ are represented as ensuing from (\ref{def:funrep}), satisfies all the defining relations. The representation is still multiplicative, but with two 
different evaluation parameters, or rapidities. In particular, one of them is boosted with respect to the 
other of an amount equal 
to the eigenvalue of one the central charges\footnote{We also notice that, as in \cite{BYang}, there is an automorphism of this realization which consists of a common constant shift of the two evaluation parameters (``boost''). Nevertheless, for the same reason explained in \cite{BYang}, the R-matrix does not depend only on the difference of the two spectral parameters, since the latter are also mixed with the representation labels of the algebra generators.}. This is particularily interesting as $C$ can take continous values and has the physical interpretation of an energy eigenvalue. Rewriting the rapidity in terms of the momentum $p$ and the energy $C$, $igu = \omega_1 = i\cot(p/2)C $, we get $\omega_2 = i\frac{e^{ip/2}}{\sin(p/2)}C$. Interestingly, the roots for which this shift occurs are precisely those involved in the modified Serre relations $\acomm{\xi^+_{2,m}}{\xi^+_{3,n}} = \gen{K}_{m+n}$, $\acomm{\xi^-_{2,m}}{\xi^-_{3,n}} = \gen{P}_{m+n}$. 

We also notice that all the three spectral parameters have the same classical limit, 
since when $g$ tends to infinity they all coincide with $ig u$. This correctly 
reproduces
the classical Yangian of \cite{BS}, as it should, since we started from the evaluation representation of \cite{BYang}\footnote{It would be interesting to repeat the calculation for the supercharges of the type found in the last paragraph of \cite{MMT}, corresponding to the classical analysis of \cite{MT}.}. 

For completeness, we report here the Drinfeld currents relative to the 
representation we have found, and their commutation relations. We define 
in the usual way such currents as
\begin{eqnarray}
\label{curr1txt}
&&\gen{E}_i^{+}(u)=\sum_{n \ge 0} \xi^+_{i,n} u^{-n-1},~~~~~~~~
\gen{F}_i^{+}(u)=\sum_{n \ge 0} \xi^-_{i,n} u^{-n-1}, \nonumber\\
&&\gen{H}_i^{+}(u)=1+ \sum_{n \ge 0} \kappa_{i,n} u^{-n-1}.
\end{eqnarray}
The evaluation of the sums is particularly easy in a multiplicative 
representation such as (\ref{mult1txt}), (\ref{mult2txt}), 
and the commutation relations can be cast in particular into 
the standard form
(cfr. \cite{YangianDouble})
\begin{eqnarray}
\label{curr2txt}
&&[\gen{H}_i^{+}(u), \gen{H}_j^{+}(v)]=0,\nonumber \\
&&\{\gen{E}_i^{+}(u),\gen{F}_j^{+}(v)\}=- \frac{\delta_{ij}}{u-v} \, [\gen{H}_i^{+}(u)-\gen{H}_j^{+}(v)],\nonumber \\
&&[\gen{H}_i^{+}(u), \gen{E}_j^{+}(v)]=- \frac{1}{2} \frac{a_{ij}}{u-v} \, [\gen{H}_i^{+}(u)(\gen{E}_j^{+}(u) - \gen{E}_j^{+}(v)) + (\gen{E}_j^{+}(u) - \gen{E}_j^{+}(v))\gen{H}_i^{+}(u)],  \nonumber  \\
&&[\gen{H}_i^{+}(u), \gen{F}_j^{+}(v)]=\frac{1}{2} \frac{a_{ij}}{u-v} \, [\gen{H}_i^{+}(u)(\gen{F}_j^{+}(u) - \gen{F}_j^{+}(v)) + (\gen{F}_j^{+}(u) - \gen{F}_j^{+}(v))\gen{H}_i^{+}(u)],  \nonumber  \\
&&\{\gen{E}_i^{+}(u),\gen{E}_j^{+}(v)\} - \{\gen{E}_j^{+}(u), \gen{E}_i^{+}(v)\}=-\frac{1}{2} \frac{a_{ij}}{u-v} [(\gen{E}_i^{+}(u) - \gen{E}_i^{+}(v))(\gen{E}_j^{+}(u) - \gen{E}_j^{+}(v)) \nonumber\\
&&\qquad \qquad \qquad \qquad \qquad \qquad - (\gen{E}_j^{+}(u) - \gen{E}_j^{+}(v))(\gen{E}_i^{+}(u) - \gen{E}_i^{+}(v))],   \nonumber \\
&&\{\gen{F}_i^{+}(u), \gen{F}_j^{+}(v)\} - \{\gen{F}_j^{+}(u), \gen{F}_i^{+}(v)\}=\frac{1}{2} \frac{a_{ij}}{u-v} [(\gen{F}_i^{+}(u) - \gen{F}_i^{+}(v))(\gen{F}_j^{+}(u) - \gen{F}_j^{+}(v)) \nonumber\\
&&\qquad \qquad \qquad \qquad \qquad \qquad - (\gen{F}_j^{+}(u) - \gen{F}_j^{+}(v))(\gen{F}_i^{+}(u) - \gen{F}_i^{+}(v))].
\end{eqnarray}

\section{Conclusions}
In this paper we have constructed the Chevalley-Serre (or Drinfeld's 
second) realization of the (centrally extended) $\alg{su(2|2)}$
Yangian symmetry underlying the AdS/CFT S-matrix, by performing a 
map on the generators obtained by \cite{BYang}. This is traditionally
the suitable basis from where to start the derivation of 
the universal R-matrix, 
and for constructing the Yangian representation theory. We have derived the 
relevant evaluation representation, and found that
the central extensions entail a peculiar feature, namely the coexistence of 
two different evaluation parameters (rapidities) for different roots. The two parameters 
are related to 
each other by a shift (boost) equal to the eigenvalue of one of the central charges, 
namely $\fC$.
We have checked this 
structure against the Serre relations, and given the realization in terms of
Drinfeld currents, which are connected to the so-called ``free-field'' 
realizations (see for example \cite{free}), and to the Faddeev-Zamolodchikov 
algebra \cite{Gleb}. 
We have also provided the Hopf algebra coproducts, and 
verified that they are algebra-homomorphisms. They satisfy a triangular 
decomposition, which is essential for the definition of a quantum double 
\cite{Drin,KT}. As a remark, we notice that the presence of two different 
evaluation parameters can complicate the derivation of our algebra from some
analog of the quantum affine version (see also \cite{NP,HSTY}).

The next important step 
will be to develop the associated representation theory, and attempt 
a construction of the universal R-matrix. 
Even though the framework we have reached is very close to the
standard one \cite{Dsecond,KT}, the new features make the direct 
application of the 
traditional procedures quite harder. On the other hand, it is known 
\cite{MMT,BS} that one additional symmetry has to play a crucial role,
and it is still unclear how to linearly embed it  
into the relations we have derived in this paper. It is also known that
at the classical level the additional central charges $\gen{P}_n, \gen{K}_n$
can be hidden in the levels of the loop algebra, so it is important to check if 
this works consistently at the level of the Yangian, as otherwise the 
quantum double would necessarily involve the undesired $\alg{sl}(2)$ automorphisms.
We reserve to come back to these issues in a future work.

\section*{Acknowledgments}
A.T. is indebted to T. Klose for the benefit derived from several discussions
and useful exchanges, and for essential help with \texttt{Mathematica}. The authors are also 
grateful to 
G. Arutyunov, A. Bassetto, N. Beisert, I. Cherednik, P. Etingof, D. Fioravanti, S. Frolov, L. Griguolo, P. Koroteev,
T. McLoughlin, J. Minahan, A. Mikhailov, A. Molev, F. Ravanini, N. Reshetikhin, S. Schaefer-Nameki, B. Stefanski, V. Stukopin, A. Tseytlin, H. Yamane and B. Zwiebel for useful discussions and comments on the manuscript.
A.T. also thanks the theory 
group of Bologna, and N. MacKay, N. Dorey and the INI in Cambridge, UK. 
This work is supported in part by funds provided by the U.S. 
Department of Energy (D.O.E.) under cooperative research agreement
DE-FG02-05ER41360. A.T. thanks Istituto Nazionale di Fisica Nucleare
(I.N.F.N.) for supporting him through a ``Bruno Rossi'' postdoctoral 
fellowship. F.S. would like to thank the Deutsche Telekom Stiftung for providing him with a PhD fellowship.

\appendix

\section{Derivation of the isomorphism}

In Section \ref{ssec:d2} we have given the isomorphism between the two realizations of the Yangian, where the main ingedients, the special elements (\ref{suspecial}), were shown to be almost the standard ones for simple Lie algebras, with a slight modification due to the appearance of the central charges in $\psucentral$. We will now derive these special elements from a more general perspective, by only demanding compatibility of the defining relations for the realizations with the appropriate coalgebra definitions, and requiring the coproduct to respect the triangular decomposition of the underlying Lie algebra. 

The Yangian generators discovered in \cite{BYang} have a multiplicative fundamental 
representation $\hat{\fJ}=ig u \fJ$, where $\fJ$ is any generator of the centrally extended $\alg{su(2|2)}$ Lie superalgebra, $g$ is the coupling constant, 
and
$u$ depends on the parameters of the representation. They are presented in the philosophy of Drinfeld's first realization, and have to be supplemented with 
an additional $\alg{u(2|2)}$-type Yangian symmetry \cite{MMT,BS} 
(see the Introduction).   
Taking as a starting point this first realization described in the main text (cfr. Section \ref{ssec:sucentral}), 
we
perform a general map of the form
\begin{align}
\label{mappa1app}
&\kappa_{i,0}=\gen{H}_i,\quad \xi^+_{i,0}=\gen{E}_i,\quad \xi^-_{i,0}=\gen{F}_i,\nonumber\\
&\kappa_{i,1}=\hat{\gen{H}}_i-v_i,\quad \xi^+_{i,1}=\hat{\gen{E}}_i-w_i,\quad \xi^-_{i,1}=\hat{\gen{F}}_i-z_i,
\end{align}
\begin{align}
\label{mappa2}
&v_i = v^{bb}_i(r,s) \fB_r \fB_s + v^{ff}_i(r,s) \fF_r \fF_s,\nonumber\\
&w_i = w^{bf}_i(r,s) \fB_r \fF_s + w^{fb}_i(r,s) \fF_r \fB_s,\nonumber\\ 
&z_i = z^{bf}_i(r,s) \fB_r \fF_s + z^{fb}_i(r,s) \fF_r \fB_s, 
\end{align}
where $\fB_r$ represents any of the ten bosonic generators, given 
(in this order) by the list
$\{\fR^1{}_1, \fR^2{}_1, \fR^1{}_2, \fL^1{}_1, \fL^2{}_1, \fL^1{}_2, \fC, \fP, \fK, 1\}$, and $\fF_s$ any of the eight fermionic ones, given (in this order) by $\{\fQ^1{}_1, \fQ^2{}_1, \fQ^1{}_2, \fQ^2{}_2, \fS^1{}_1, \fS^2{}_1, \fS^1{}_2, \fS^2{}_2\}$. The presence of $1$ among the bosonic elements allows in principle for linear (and constant) terms in the map.
 We then require the new generators to satisfy the desired relations (\ref{relazionizero}) 
\begin{align}
\label{relazionizeroapp}
&[\kappa_{i,m},\kappa_{j,n}]=0,\quad [\kappa_{i,0},\xi^+_{j,m}]=a_{ij} \,\xi^+_{j,m},\nonumber\\
&[\kappa_{i,0},\xi^-_{j,m}]=- a_{ij} \,\xi^-_{j,m},\quad \{\xi^+_{i,m},\xi^-_{j,n}\}=\delta_{i,j}\, \kappa_{j,n+m},\nonumber\\
&[\kappa_{i,m+1},\xi^+_{j,n}]-[\kappa_{i,m},\xi^+_{j,n+1}] = \frac{1}{2} a_{ij} \{\kappa_{i,m},\xi^+_{j,n}\},\nonumber\\
&[\kappa_{i,m+1},\xi^-_{j,n}]-[\kappa_{i,m},\xi^-_{j,n+1}] = - \frac{1}{2} a_{ij} \{\kappa_{i,m},\xi^-_{j,n}\},\nonumber\\
&\{\xi^+_{i,m+1},\xi^+_{j,n}\}-\{\xi^+_{i,m},\xi^+_{j,n+1}\} = \frac{1}{2} a_{ij} [\xi^+_{i,m},\xi^+_{j,n}],\nonumber\\
&\{\xi^-_{i,m+1},\xi^-_{j,n}\}-\{\xi^-_{i,m},\xi^-_{j,n+1}\} = - \frac{1}{2} a_{ij} [\xi^-_{i,m},\xi^-_{j,n}],
\end{align}
together with suitable Serre relations, which we leave unspecified for the 
moment. 

\subsection{Constraints on the coefficients}
We will impose all of the above relations to be satisfied, using the specific representation of \cite{BYang}, for {\it arbitrary parameters\footnote{\rm The set of solutions we find is therefore a subset of the solutions one would obtain imposing the actual dependence of $u$ on the parameters of the representation.}}
$a$, $b$, $c$ and $u$, and constant  
coefficients of the map (i.e. independent on the representation labels). We will comment later on the issue of promoting the resulting 
relations at a universal level.
This operation puts a list of constraints (which we shall not report here) 
on the coefficients, but at the same times leaves 
many of them arbitrary\footnote{The map (\ref{mappa1app}), (\ref{mappa2}), in its generality, 
is however partly redundant in the fundamental representation.}.
This means that there are in principle many ways to achieve the desired 
realization. However, this is not all one needs to impose, in order to have a 
consistent Hopf algebra, since the coproducts of these generators must also 
satisfy the above relations. Such coproduct is directly obtained from the map, by using the knowledge of the original coalgebra structure \cite{GH,PST}, and the homomorphism property. Namely one has

\begin{align}
&\Delta(\kappa_{i,0})=\Delta(\gen{H}_i),\quad \Delta(\xi^+_{i,0})=\Delta(\gen{E}_i),\quad \Delta(\xi^-_{i,0})=\Delta(\gen{F}_i),\nonumber\\
&\Delta(\kappa_{i,1})=\Delta(\hat{\gen{H}}_i)-\Delta(v_i),\quad \Delta(\xi^+_{i,1})=\Delta(\hat{\gen{E}}_i)-\Delta(w_i),\quad \Delta(\xi^-_{i,1})=\Delta(\hat{\gen{F}}_i)-\Delta(z_i),
\end{align}
\begin{align}
&\Delta(v_i) = v^{bb}_i(r,s) \Delta(\fB_r) \Delta(\fB_s) + v^{ff}_i(r,s) \Delta(\fF_r) \Delta(\fF_s),\nonumber\\
&\Delta(w_i) = w^{bf}_i(r,s) \Delta(\fB_r) \Delta(\fF_s) + w^{fb}_i(r,s) \Delta(\fF_r) \Delta(\fB_s),\nonumber\\ 
&\Delta(z_i) = z^{bf}_i(r,s) \Delta(\fB_r) \Delta(\fF_s) + z^{fb}_i(r,s) \Delta(\fF_r) \Delta(\fB_s). 
\end{align}
We use the conventions in \cite{BYang} for the coproducts of the original generators, with braiding factor $\fU$. In our basis, this is generated by
\begin{eqnarray}
\label{Hopforig}
&&\Delta(\xi^\pm_{1,0}) = \xi^\pm_{1,0} \otimes 1 + \fU^\pm \otimes \xi^\pm_{1,0},\nonumber\\
&&\Delta(\xi^\pm_{j,0}) = \xi^\pm_{j,0} \otimes 1 + \fU^\mp \otimes \xi^\pm_{j,0},
\end{eqnarray}
with $j=2,3$, for the Lie algebra generators, and by the corresponding 
first level Yangian formulas of \cite{BYang}, see Table \ref{tab:drinf1co}.
 
As a first step, we require an appropriate triangular decomposition in the spirit of the treatment in 
\cite{KT}. We will ask the coproducts to be schematically of the form
\begin{align}
\label{decozeroapp}
&\Delta(\xi^+_{i,1}) =  \xi^+_{i,1} \otimes 1 + \fU^{[i]} \otimes \xi^+_{i,1} + \xi^+_{i,0} \otimes \kappa_{i,0} + \alg{E} \fU^{[Y]}\otimes Y,\nonumber\\
&\Delta(\xi^-_{i,1}) = \xi^-_{i,1} \otimes 1 + \fU^{-[i]} \otimes 1 + \kappa_{i,0} \fU^{-[i]} \otimes \xi^-_{i,0} + Y \otimes \alg{F},\nonumber\\
&\Delta(\kappa_{i,1}) = \kappa_{i,1} \otimes 1 + 1 \otimes \kappa_{i,1} + \kappa_{i,0} \otimes \kappa_{i,0} + \alg{E} \fU^{[\alg{F}]}  \otimes \alg{F}.
\end{align}
We refer to the comments in the text (cfr. formula (\ref{decozero2})) 
for the meaning of the notation in (\ref{decozeroapp}).

In order to achieve such a decomposition, it is enough to impose the following additional constraints: 
\begin{align}
&v^{bb}_3(4,7) = - 1 - v^{bb}_3(7,4),\quad v^{bb}_3(4,8) = - v^{bb}_3(8,4),\quad v^{bb}_3(4,9) = - v^{bb}_3(9,4).
\end{align}
We afterwards impose that all the coproducts obtained in this way also satisfy the above relations (\ref{relazionizeroapp}){\footnote{We will again impose
physical constraints only at the very end, for the simplicity of the calculation. The set of solutions is therefore a subset of the ones we would obtain otherwise.}. This qualifies the coproducts as algebra homomorphisms. One obtains in this way additional constraints on the coefficients, in particular, the vanishing of the constant piece of the map, $v^{bb}_i(10,10)=0$.

At this stage, one can check that the antipode S, similarly obtained from the original one using the anti-homomorphism property
\begin{eqnarray}
S(xy) = (-)^{|x||y|} S(y) S(x), 
\end{eqnarray}
preserves the relations (\ref{relazionizeroapp}), and is automatically consistent with the charge conjugation rule \cite{J,PST,Gleb} 
\begin{align}
\label{antipoderel}
{\rm S}(\fJ_n(x^\pm))
={{\cal{C}}}^{-1}\bigl[\fJ_n(1/x^\pm)\bigr]^{st}{{\cal{C}}},
\end{align}
in the fundamental representation, 
with one and the same 
charge conjugation matrix ${\cal{C}}$. Explicit formulas can alternatively be derived from the final expression for the coproducts (see below), using the defining Hopf algebra property of the antipode $\mu\circ({\rm S}\otimes 1)\circ\Delta=\eta\circ\epsilon$
($\mu$ is the algebra multiplication, $\eta$ the unit, $\epsilon$ the counit).

\subsection{Image of the map and triangular decomposition}
Computing the image of the original generators under the map (\ref{mappa1app}), (\ref{mappa2}) gives at 
this point the following result:
\begin{align}
\label{mult}
&\kappa_{i,1} = \omega_i \kappa_{i,0},\quad \xi^+_{i,1} = \omega_i \xi^+_{i,0},\quad \xi^-_{i,1} = \omega_i \xi^-_{i,0},
\end{align}
with 
\begin{align}
&\omega_1= ig u + \frac{i}{2} [2 z^{fb}_3(3,10) + z^{bf}_3(1,3) - 2 z^{bf}_3(3,1) + z^{bf}_3(4,3) + 2 z^{bf}_3(6,4) + 2 z^{bf}_3(10,3)],\nonumber\\
&\omega_2 = \omega_3 = \omega_1 - C. 
\end{align}
We  can see that the representation is still multiplicative, but with two 
different evaluation parameters, or rapidities. In particular, one of them is boosted with respect to the 
other of an amount equal 
to the eigenvalue of one the central charges.
It is also obvious that the constant shift in $\omega_1$ is totally unessential, and we impose a further constraint on the coefficients of the map such that this shift vanishes, ending up with just the representation reported in the main text (cfr. (\ref{mult1txt}), (\ref{mult2txt})):
\begin{align}
\label{mult2}
&\omega_1= ig u,\nonumber\\
&\omega_2 = \omega_3 = ig u - C. 
\end{align}
By making recursive use of the defining relations (\ref{relazionizeroapp}), one can
easily generate all higher levels, obtaining an evaluation representation of
the form
\begin{eqnarray}
\label{mult1}
&\kappa_{i,n} = {\omega_i}^n \kappa_{i,0},\quad \xi^+_{i,n} = {\omega_i}^n \, \xi^+_{i,0},\quad \xi^-_{i,n} = {\omega_i}^n \, \xi^-_{i,0}.
\end{eqnarray}

Since, after making use of all the constraints described until now, 
many coefficients of the map (\ref{mappa1app}), (\ref{mappa2}) are still undetermined, we will perform
a final choice, which puts the image generators and their coproduct 
in the 
simplest and most symmetric form. The triangular decomposition
advertised in (\ref{decozeroapp}) is then mostly evident, 
and the map converges to almost the original
Drinfeld prescription \cite{Dsecond}, reproducing the result reported in the main text. This amounts to setting to zero many 
of the coefficients, and fixing the remaining ones such as to produce precisely the special elements (\ref{suspecial}) as well as the coproducts (\ref{coplevonetxt}).

One can check the triangular decomposition according to the subdivision
 in positive and negative roots ensuing from Section \ref{ssec:sucentral}. In particular, the
non-simple roots $\fR^1{}_2$, $\fL^2{}_1$ and $\fS^1{}_1$  
are positive,
while  $\fR^2{}_1$, $\fL^1{}_2$ and $\fQ^1{}_1$ are negative. Furthermore,
$\fK$ is generated inside $I_+$, and $\fP$ inside $I_-$. 

Let us now turn to the issue of the Serre relations. The representation (\ref{mult1}), (\ref{mult2}) is multiplicative, and, in particular, it has the {\it same} evaluation parameter precisely for the simple roots with indices $2$ and $3$, which allows to define unambiguously the higher central charges that have to appear in the Serre relations. We can 
immediately verify that the following relations are in fact satisfied 
\begin{eqnarray}
&&i\neq j, \, \, \, \, \, n_{ij}=1+|a_{ij}|,\, \, \, \, \, Sym_{\{k\}} [\xi^+_{i,k_1},[\xi^+_{i,k_2},\dots \{\xi^+_{i,k_{n_{ij}}}, \xi^+_{j,l}\}\dots\}\}=0,\nonumber\\
&&i\neq j, \, \, \, \, \, n_{ij}=1+|a_{ij}|,\, \, \, \, \, Sym_{\{k\}} [\xi^-_{i,k_1},[\xi^-_{i,k_2},\dots \{\xi^-_{i,k_{n_{ij}}}, \xi^-_{j,l}\}\dots\}\}=0,\nonumber\\
&&\text{except for} \, \, \, \, \, \, \, \, \, \{\xi^+_{2,n},\xi^+_{3,m}\} = \alg{K}_{n+m}, \qquad  \{\xi^-_{2,n},\xi^-_{3,m}\} = \alg{P}_{n+m}, 
\end{eqnarray}
where $\alg{K}_{n} = {\omega_2}^n \alg{K}$ and $\alg{P}_{n} = {\omega_2}^n 
\alg{P}$ are central elements. 
Notice that this occurrence represents a quite non-trivial consistency check 
of the realization we have found.

Once we have reached this result, we can come back to the issue of
``universality'' of these relations. If it is true that the single steps 
where performed in the specific representation of \cite{BYang}, the 
structure reported in Section \ref{summary} 
seems to be consistent by itself, and there is no apparent 
obstacle in attributing a universal meaning to it. One may take 
it as a starting point for the subsequent analysis of the universal R-matrix,
making use of the new evaluation representation (\ref{mult1}), (\ref{mult2}), 
since there seems
to be no {\it a priori} ways to discriminate on its abstract validity.

\end{document}